\documentclass[final,english]{bullsrsl}[2022/06/15]

\usepackage[latin1]{inputenc}
\usepackage[T1]{fontenc}

\usepackage{natbib} 
\usepackage{graphicx}
\usepackage[dvipsnames]{xcolor}
\usepackage{amsmath}
\usepackage{siunitx}

\newcommand{\teff}{$T_\textrm{eff}$}
\newcommand{\logg}{$\log g$}

\newcommand{\kms}{km\,s$^{-1}$}

\newcommand{\vrot}{$v_{\rm rot}$}
\newcommand{\halpha}{H$\alpha$}
\newcommand{\hbeta}{H$\beta$}
\newcommand{\hgamma}{H$\gamma$}

\newcommand{\dsct}{$\delta$\,Scuti}

\begin{document}
\title{Curve of Growth Analysis of SZ\,Lyn\\ 
}

\author[affil={1},corresponding]{Janaka}{Adassuriya}
\author[affil={2}]{Shashikiran}{Ganesh}
\author[affil={3}]{Peter}{De Cat}
\author[affil={4}]{Santosh}{Joshi}
\author[affil={1}]{Chandana}{Jayaratne}

\affiliation[1]{Astronomy and Space Science Unit, Department of Physics, University of Colombo, Sri Lanka}
\affiliation[2]{Astronomy and Astrophysics Division, Physical Research Laboratory, Ahmedabad, India}
\affiliation[3]{Royal Observatory of Belgium, Ringlaan 3, B-1180 Brussels, Belgium}
\correspondance{janaka@phys.cmb.ac.lk}
\affiliation[4]{Aryabhatta Research Institute of Observational Science (ARIES), Manora Peak-Nainital, India}

\date{13th October 2020}
\maketitle


%

\begin{abstract}
We present one high-resolution and a time series of 561 low-resolution follow-up spectroscopic observations of SZ\,Lyn. 
It is a high-amplitude \dsct-type pulsating star in a binary system. 
The photometric observations reveal the existence of radial and non-radial oscillation modes in SZ\,Lyn. In spectroscopy, the variation of equivalent width of the line profiles reflects the temperature variations. The equivalent widths of the Balmer lines, \halpha, \hbeta, and \hgamma\ were measured over the pulsation cycle of SZ\,Lyn using time sequence spectra. Hence, the temperature profile of SZ\,Lyn was derived using the curve of growth analysis. Furthermore, the stellar parameters were determined through the best fit analysis of observed and synthetic high-resolution spectral lines. The best fit determines a model of \teff\,=\,6750\,K, \logg\,=\,3.5\,dex, and \vrot\,=\,10\,\kms\ for solar abundance. 
\end{abstract}


\section{Introduction}
SZ\,Lyn (HD\,67390) is a \dsct-type pulsating star in a binary system. 
This binary system is characterized as a single-lined spectroscopic binary \citep{Gazeas2004} where the brighter component is SZ\,Lyn 
with a dominant radial mode at 8.296 d$^{-1}$ simultaneously excited with several non-radial p-modes and a possible g-mode \citep{adassuriya2021asteroseismology}.The observed minus calculated (O-C) diagram redefined the binary orbital parameters, resulting in an orbital period of 1187$\pm$15 days, a projected semi-major axis $a sin(i)$ of 1.4$\pm$0.1$\times$10$^{8}$\,km, and an eccentricity of 0.18$\pm$0.07 using 378 light maxima \citep{adassuriyalight}.

Spectroscopic observations appropriate for a pulsation analysis of short-period \dsct\ variable stars are hardly found in the literature. One good spectrum is already sufficient to determine the basic stellar parameters of a star (e.g. \citealt{niemczura2015spectroscopic}). Most spectroscopic surveys are carried out for this purpose. However, time-resolved spectra with a high resolution (R$>$40000) and a sufficiently high signal-to-noise ratio (SNR$>$200) are needed to resolve and identify radial and non-radial pulsation modes with very low amplitudes (e.g. \citealt{Zima,zima2008spectroscopic,aerts2008spectroscopic,2010aste.book.....A}). \citet{bardin1984orbital} and \citet{garbuzov1983spectroscopic} observed SZ\,Lyn spectroscopically to study the binarity and pulsation. They determined the orbital parameters. These are the only spectroscopic studies of SZ\,Lyn available to date. In this paper, we present a time series of low-resolution spectroscopic observations of SZ\,Lyn that we use for a curve of growth analysis in terms of excitation temperature.

\section{Observations in orbital and pulsation phases}

\begin{figure}
\centering
\includegraphics[scale=0.475]{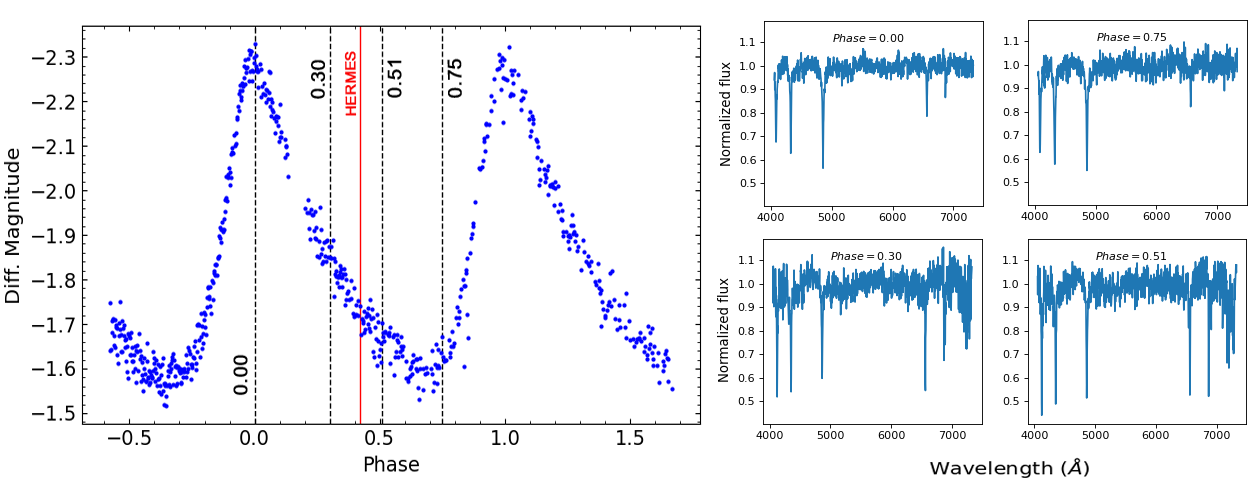}
\bigskip
\begin{minipage}{12cm}
\caption{Left: Pulsation phase of the B-band light curve of SZ\,Lyn. The phase conversion is based on the reference epoch 2456664.261713 HJD and a pulsation period of 0.120526 days \citep{adassuriya2021asteroseismology}. Right: Low-resolution spectra of four selected phases observed by LISA at Mount Abu observatory. The corresponding pulsation phases are given on top of each panel and indicated as vertical dashed lines in the left panel.}
\label{fig:1}
\end{minipage}
\end{figure}

A series of 561 low-resolution (R$\approx$1000) spectra covering the entire visible range of SZ\,Lyn were obtained from Mount Abu InfraRed Observatory (MIRO) from December 8 to 12, 2016 (2457731-2457735 JD). Observations were made with the Corrected Dall-Kirkham (CDK) 50 cm, f/6.8 equatorial mount telescope equipped with the LISA spectrograph. Details about the observational setup are provided by \citet{Ganesh2013}. 
The effective time base of the Mount Abu observations is 9.24 hours, covering the full pulsation period of 2.88\,hours of SZ\,Lyn. 
A high resolution (R$\approx$85000) high signal-to-noise (SNR$>$150) spectrum of SZ\,Lyn was acquired on September 17, 2020 (2459109.743 JD) 
with the High Efficiency and Resolution Mercator Echelle Spectrograph (HERMES; \citealt{Raskin2011A&A...526A..69R}) attached to the 1.2-m Mercator telescope at the Roque de los Muchachos Observatory on the Canary Island La Palma (Spain) at an altitude of 2333 meters.

The initial data reduction steps including bias, flat-field, wavelength calibration, normalization, and heliocentric velocity correction were done for all the 561 spectra using IRAF. 
The orbital velocity correction due to binary is subsequently applied to the spectra by calculating the orbital phase of 
each observation epoch. \citet{mcnamara1976radial} mentioned the orbital phase of SZ\,Lyn to be 0.78 at 2442473.748 HJD. This leads to predicting the epoch of the primary eclipse with the orbital period of 1187 days and yields the primary eclipse at 2442733.788 HJD. The primary eclipse of 2445156.600 HJD was observed by \citet{bardin1984orbital} and is also matched with McNamra's determination of primary epoch of 2442733.788 HJD. Therefore, we used 2445156.600 HJD to determine the orbital phase of the Mount Abu and Mercator 
observations, resulting in an orbital phase of 0.688 and 0.800, respectively.
Hence, both orbital phases result in the red-shifted spectra. These red shifts were estimated with the average orbital velocity of SZ Lyn of 34 \kms\ \citep{bardin1984orbital} for H$\alpha$ (6563 \AA), H$\beta$ (4861 \AA) and H$\gamma$ (4340 \AA) lines. Hence, the red shifts of Mount Abu and Mercator observations were determined as 0.61\,\AA\ and 0.67\,\AA\ respectively. These shifts were applied to the observed data and were confirmed by a comparison with the synthetic spectra. 

Similarly, the pulsation phases of the HERMES and Mount Abu spectra were calculated with reference to the light curve maxima at 2456664.261 HJD \citep{adassuriya2021asteroseismology}. The pulsation phases corresponding to the HERMES spectrum and the Mount Abu spectra are indicated as vertical lines on the light curve of SZ\,Lyn in the left panel of Fig.\,\ref{fig:1}.
The two spectra at phases 0.00 and 0.75 of the right panel of Fig.\,\ref{fig:1} show more stable line profiles over the low noise continuum while at phases 0.30 and 0.51 there is comparatively higher noise in the continuum. From the left panel of Fig.\,\ref{fig:1}, it is clear that 0.00 and 0.75 phases are at the peak of the magnitude and on the rising edge of the magnitude, respectively. This indicates that in those two phases, the star begins brightening steadily and rising the magnitude more smoothly than in the other two phases of 0.30 and 0.51. More details of this analysis will be given in our forthcoming paper.

\section{Synthetic spectra and stellar parameters}

\begin{figure}
\centering
\includegraphics[scale=0.3]{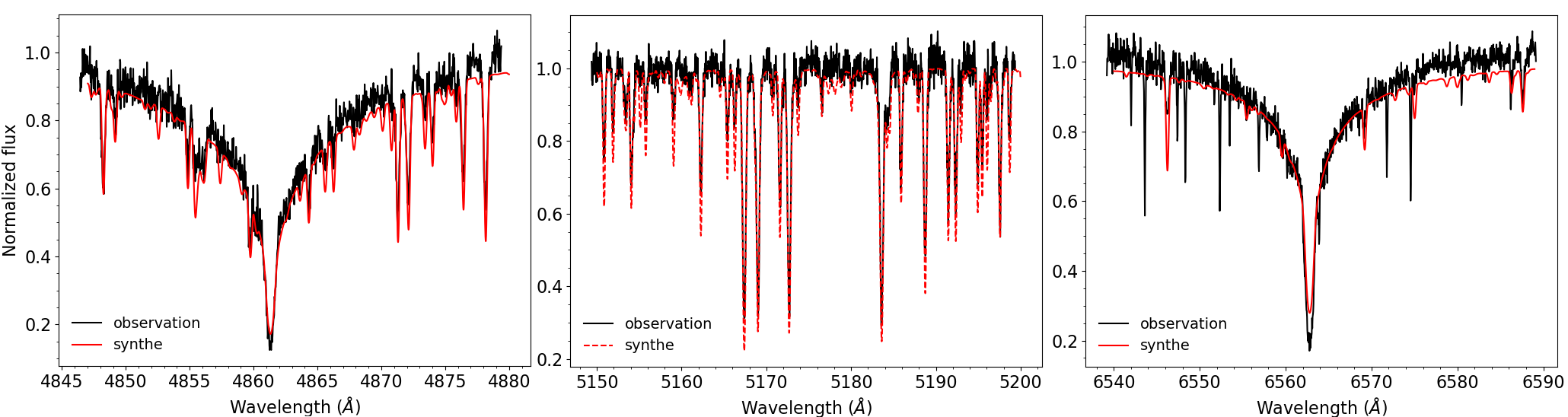}
\bigskip
\begin{minipage}{12cm}
\caption{High-resolution line profiles of SZ\,Lyn. The black lines are observations and red represents the synthetic line profiles produced by SYNTHE. From left to right, \hbeta, Mg(I) triplet and \halpha\ respectively. The \hbeta\ and MgI triplet were approximated by a model with \teff\,=\,6750\,K, \logg\,=\,3.5\,dex and \vrot\,=\,10\,\kms\ and \halpha\ approximated to slightly different model of \teff\,=\,6500\,K, \logg\,=\,4.0\,dex and \vrot\,=\,10\,\kms.}
\label{fig:2}
\end{minipage}
\end{figure}

High-resolution and low-resolution synthetic spectra were computed using model atmospheres produced by the ATLAS 12 and ATLAS 9 codes, respectively \citep{1993KurCD,castelli2005atlas12}. The model atmospheres were fed to the SYNTHE code \citep{1993Kurucz} to produce the synthetic spectra. All spectra were generated through opacity distribution functions (ODFs) and for solar abundances from \citet{anders1989abundances}. The \dsct\ stars are well characterized by a microturbulence of 2\,kms$^{-1}$\ and a high mixing length parameter of typically 1.8 - 2.0 \citep{bowman2018mnras}. 
We used the highest mixing length 1.25 in ATLAS 9 models. The instrumental broadening of the synthetic spectra can be approximated by Gaussian, $\text{Sin(x)/x}$ and rectangular profiles with \kms\ or full width at half maxima (FWHM) scales. We used a Gaussian profile of FWHM at 48000 resolution scale \citep{castelli2004spectroscopic}. The parameters, temperature (\teff), gravity (\logg) and rotation (\vrot) were kept as variables throughout the model fitting. 
In Fig.\,\ref{fig:2}, some of the wavelength regions of the observed high-resolution HERMES spectrum of SZ\,Lyn are compared with the best fitting solar abundance models, corresponding to \teff\,=\,6750\,K, \logg\,=\,3.5\,dex, and \vrot\,=\,10\,\kms\ for the \hbeta\ line (left) and the Mg{\sc i} triplet (middle) and to \teff\,=\,6500\,K, \logg\,=\,4.0\,dex and \vrot\,=\,10\,\kms\ the \halpha\ line (right).
Due to the variability of the star, the spectral range changes from A to F. The HERMES spectrum was obtained at the fading phase of 0.42 in pulsation where the observed magnitude is highly scattered (Fig.\,\ref{fig:1}). This decline phase can be identified in equivalent widths of \hgamma, \hbeta\ and \halpha\ in Fig.\,\ref{fig:3}. Therefore, the HERMES spectrum was taken at the diminishing phase of the transition. Due to this instability, the blue and the red wavelength regions behave differently where \hbeta\ and \halpha\ converge to slightly different models. Nevertheless, the blue side is more stable and both \hbeta\ and Mg{\sc i} lines are perfectly matched. Hence, \teff\,=\,6750\,K, \logg\,=\,3.5\,dex was taken as the average model to represent SZ\,Lyn.

\section{Curve of Growth and Excitation Temperature}

\begin{figure}
\centering
\includegraphics[scale=0.6]{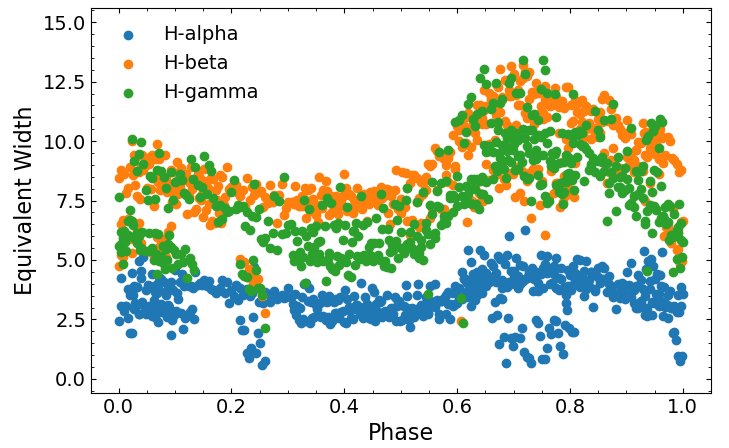}
\bigskip
\begin{minipage}{12cm}
\caption{Distribution of the equivalent width of Balmer lines of SZ\,Lyn. The phase calculation is based on T$_0$=2456664.261713 HJD, the maximum magnitude observed on 6$^{th}$ January 2014 photometric observation at Mount Abu observatory.}    
\label{fig:3}
\end{minipage}
\end{figure}

\begin{figure}
\centering
\includegraphics[scale=0.5]{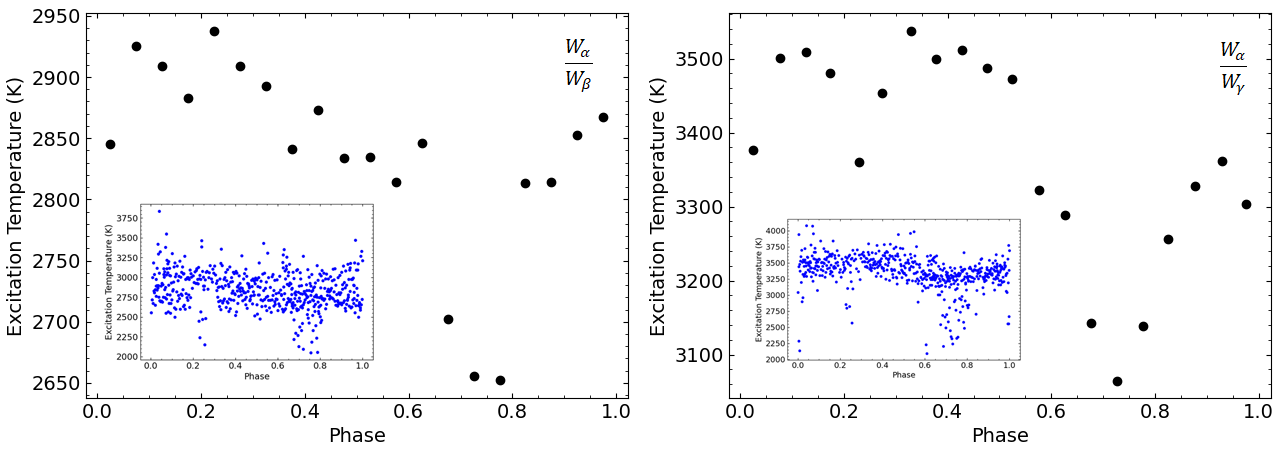}
\bigskip
\begin{minipage}{12cm}
\caption{The excitation temperature of the lines \halpha\ and \hbeta\ (left) and the lines \halpha\ and \hgamma\ (right) over the pulsation cycle of SZ\,Lyn. The observations were binned in 0.05 phases to reduce the noise. The inset shows the excitation temperature of all 561 spectra.}    
\label{fig:4}
\end{minipage}
\end{figure}

The spectroscopic surveys and follow-up observations are more common in terms of the determination of stellar global parameters 
(e.g. \citealt{niemczura2015spectroscopic,fleming2015apogee,johnson2017california}).
This is the first time that a high number of high-cadence
spectroscopic observations were carried out for SZ Lyn in terms of pulsation analysis. We used a total of 561 low-resolution spectra in the pulsation analysis of SZ\,Lyn. The equivalent width ($W_\lambda$) of the absorption lines indicates the emission of energy and hence the temperature variation of pulsating stars. Therefore, the equivalent widths ($W_\lambda$) of the Hydrogen lines (\hgamma, \hbeta, \halpha) were determined for all the 561 spectra of SZ\,Lyn as shown in Fig.\,\ref{fig:3}. 

The curve of growth describes how 
$W_\lambda$
depends on the number of absorbing atoms ($N$). For optically thin lines, 
$W_\lambda$ is proportional to 
$N$ times the oscillator strength ($f_s$) \citep{bohm1989introduction}. Therefore, the Boltzmann formula can be approximated as 

\begin{equation}
W_\lambda \propto \frac{N_0}{g_0} g_sf_s\lambda^2e^{-{\frac{\chi_s}{kT}}}
\end{equation}
where, 
$g_0$ and $g_s$ are the statistical weights of the ground and excited state, respectively, $f_s$ is the oscillator strength of the excited state \citep{green1957oscillator}, and $\chi_s$ is the energy difference between the excited and ground state. For two such lines, one with index 1 and another with index 2, the ratio of the equivalent widths gives the excitation temperature ($T_{exc}$), which is determined from the excitation difference of the two levels.

\begin{equation}
\frac{W_{\lambda1}}{W_{\lambda2}} = \frac{g_1f_1}{g_2f_2}\frac{\lambda_1^2}{\lambda_2^2} e^{-{\frac{(\chi_1-\chi_2)}{kT_{exc}}}}
\end{equation}

The ratio of the line width of $W_\alpha/W_\beta$ and $W_\alpha/W_\gamma$ were used to determine 
$T_{exc}$ over the pulsation phase of SZ\,Lyn as shown in Fig.\,\ref{fig:3}. 
It shows that both temperature variations are in agreement with the light curve variation in Fig.\,\ref{fig:1}. However, a detailed analysis should be done to see any phase difference between the magnitude and temperature variations. This is expected to be addressed in our forthcoming paper. 
Moreover, we can conclude that the variation in $T_{exc}$ is smoother
in $W_\alpha/W_\gamma$ than $W_\alpha/W_\beta$ by looking at the inset plots in Fig.\,\ref{fig:4}. Furthermore, the calculation of $T_{exc}$ from the ratio of $W_\beta/W_\gamma$ is found to be very noisy. This leads to 
the conclusion that selecting profiles with larger wavelength differences
would give more clear excitation temperature profiles. 

\section{Conclusion}

The stellar parameters of SZ\,Lyn were determined using the observed and synthetic spectra. The best overall model corresponds to
\teff\,=\,6750\,K, \logg\,=\,3.5\,dex and \vrot\,=\,10\,\kms\ for a solar abundance star. The variations of the equivalent width of the time sequence spectra were used to determine the profile of excitation temperature.
They are found to be in phase with the observed magnitude variations of SZ\,Lyn.  

\begin{acknowledgments}
JA acknowledges the local support given by the organizing committee of the third BINA workshop under the project approved by the International Division, Department of Science and Technology (DST, Govt. of India; DST/INT/BELG/P-09/2017) and the Belgian Federal Science Policy Office (BELSPO, Govt. of Belgium; BL/33/IN12). Work at PRL is supported by the Department of Space, Government of India.
IRAF is distributed by the National Optical Astronomy
Observatories, which is operated by the Association of Universities for Research in Astronomy, Inc. (AURA) under cooperative agreement with the National Science Foundation. 
Based on observations obtained with the HERMES spectrograph, which is supported by the Research Foundation - Flanders (FWO), Belgium, the Research Council of KU Leuven, Belgium, the Fonds National de la Recherche Scientifique (F.R.S.-FNRS), Belgium, the Royal Observatory of Belgium, the Observatoire de Gen{\`e}ve, Switzerland and the Th{\"u}ringer Landessternwarte Tautenburg, Germany.
\end{acknowledgments}

\begin{furtherinformation}

\begin{orcids}
\orcid{0000-0002-6070-956X}{Janaka}{Adassuriya}
\orcid{0000-0002-7721-3827}{Shashikiran}{Ganesh}
\orcid{0000-0001-5419-2042}{Peter}{De Cat}
\end{orcids}

\begin{authorcontributions}

This work is the result of collective efforts by all the co-authors with relevant contributions.

\end{authorcontributions}

\begin{conflictsofinterest}
The authors declare no conflict of interest.
\end{conflictsofinterest}

\end{furtherinformation}

\bibliographystyle{bullsrsl-en}

\bibliography{extra}

\end{document}